\documentclass[osajnl,twocolumn,showpacs,superscriptaddress,11pt]{revtex4-1} %% use 11pt for Applied Optics
\usepackage[draft]{hyperref} %% optional
\usepackage{xfrac}
\usepackage{multirow}
\usepackage[T1]{fontenc}
\usepackage{ae}
\usepackage{amsmath, amssymb}
\usepackage{color}
\usepackage{parskip}
\usepackage{graphicx,float}
\usepackage{caption}
\usepackage[german, english]{babel}
\usepackage{xspace}
\usepackage{tabularx}
\usepackage[dvips,ps2pdf]{thumbpdf}
\usepackage{textcomp}
\usepackage{mathtools}

\begin{document}

\title{Feedback control of optical beam spatial profiles using thermal lensing}

%\author{Zhanwei Liu,$^{1,2,}$\symbolfootnote[2]{This work was performed when these authors were based at the University of Florida}  Paul Fulda,$^{1,*}$ Muzammil A. Arain,$^{3,\dagger}$ Luke F. Williams,$^1$ Guido Mueller,$^1$ David B. Tanner,$^{1}$ David H. Reitze,$^{1,4}$}
%\address{$^1$Department of Physics, University of Florida, P.O. Box 118440,\\ Gainesville, Florida 32611, USA}
%\address{$^2$Currently: School of Applied and Engineering Physics, Cornell University\\ Ithaca, New York 14850, USA}
%\address{$^3$KLA-Tencor Company, One Technology Drive, Milpitas, California 95035, USA}
%\address{$^4$Currently: LIGO Laboratory, California Institute of Technology, MS 100-36, \\Pasadena, California 91125, USA}
%\address{$^*$pfulda@phys.ufl.edu}

\author{Zhanwei Liu$^\dagger$}
\affiliation{School of Applied and Engineering Physics, Cornell University\\ Ithaca, New York 14850, USA}

\author{Paul Fulda}\email{Corresponding author: pfulda@phys.ufl.edu \\
$^\dagger$This work was performed when these authors were based at the University of Florida}
\affiliation{Department of Physics, University of Florida, P.O. Box 118440,\\ Gainesville, Florida 32611, USA}

\author{Muzammil A. Arain$^\dagger$}
\affiliation{KLA-Tencor Company, One Technology Drive, Milpitas, California 95035, USA}

\author{Luke Williams}
\author{Guido Mueller}
\author{David B. Tanner}
\affiliation{Department of Physics, University of Florida, P.O. Box 118440,\\ Gainesville, Florida 32611, USA}

\author{David H. Reitze}
\affiliation{Department of Physics, University of Florida, P.O. Box 118440,\\ Gainesville, Florida 32611, USA}
\affiliation{Currently: LIGO Laboratory, California Institute of Technology, MS 100-36, \\Pasadena, California 91125, USA}

\begin{abstract}
A method for active control of the spatial profile of a laser beam using 
adaptive thermal lensing is described. A segmented electrical heater was used to generate thermal gradients across a transmissive 
optical element, resulting in a controllable thermal lens. The segmented heater also allows the generation of cylindrical lenses, and 
provides the capability to steer the beam in both horizontal and vertical planes. Using this device as an actuator, a feedback control loop was developed to 
stabilize the beam size and position.
\end{abstract}
\maketitle %% null function with osajnl.sty

\section{Introduction}
High laser powers in scientific experiments and industrial applications often lead to the generation 
of thermal lenses in optical systems, causing unwanted changes in spatial beam profiles. 
Absorption of high-power laser radiation in transmissive optical materials deposits heat in a non-uniform spatial 
distribution, leading to a temperature gradient and therefore refractive index gradient across the optic surface. 
The resulting thermal gradient index lens changes the spatial mode of the laser and can cause 
aberrations which lead to modal distortions \cite{Foster70, Greninger86, Hello90}. 

We report the development of an active device designed to counteract the effects of thermal lensing, first described in Ref. \cite{Arain10}. 
Four individually addressable heating elements are used to generate thermal gradients across a transmissive optical element, 
resulting in a controllable thermal lens having both spherical and (if necessary) cylindrical components, 
providing the capability to focus and steer the beam. In this article we demonstrate the use of this device as an actuator in a beam shaping 
feedback loop, operating over a wide non-linear range to correct for impulsive thermally generated beam distortions. 

While potential applications of such an adaptive device exist in areas of laser material processing \cite{Schmiedl00}, image 
processing \cite{Sato79, 	Jablonowski75}, and optical displays \cite{Kelly00}, this work is focused mainly on applications to Advanced 
LIGO (aLIGO), a high power laser interferometer which aims to detect gravitational waves from merging neutron stars, black holes and other 
astrophysical sources \cite{Harry10}. 

The aLIGO pre-stabilized laser system generates a 200\,W laser beam at 1064\,nm \cite{Winkelmann11}. This high power beam is expected to 
create thermal lenses in many of the optics comprising the detector. In turn, this thermal lensing causes a laser power dependence in the mode 
matching between the beam from the pre-stabilized laser and the various optical cavities present in the interferometer. 
In particular, the mode matching between the beam transmitted through the input mode cleaner and the main interferometer power recycling cavity 
is expected to be dependent on the laser power \cite{Arain07note}. A system which can be used to correct for the power dependence is therefore highly desirable in order to 
maintain good mode matching into the interferometer. For this application, where repositioning of lenses is not easily achieved due to the ultrahigh-vacuum 
environment, adaptive optical elements present the most practical solution for maintaining optimal beam parameters in the high laser power regime. 

The device presented here is not the first method developed 
to compensate thermal lenses; there are a number of thermal compensation techniques, using for example negative thermo-optic coefficient materials 
\cite{Michel04}, CO$_2$ laser heating and electrical heating of the optical elements \cite{Arain07,Lawrence04}, tunable liquid crystals \cite{Sato79} or 
deformable mirrors \cite{Schwartz06}. 
However, this device is vacuum compatible and high laser power compatible, as well as being versatile and reliable. 
Vacuum and high laser power compatibility are essential for use of the device in aLIGO, where it may be employed within the ultrahigh-vacuum system and 
will be required to transmit the full science laser power. 
\begin{figure}[htb]
\centering
\includegraphics[width=0.4\textwidth]{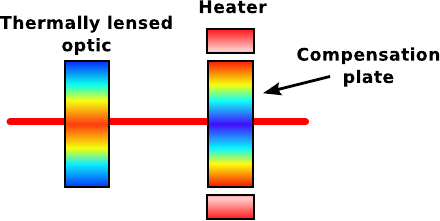}
\caption{The conceptual design of thermal compensating system. The heated compensation plate is used to compensate the 
laser induced thermal lens.}
\label{fig:conceptualdesign}
\end{figure}
\begin{figure}
\centering
\includegraphics[width=0.25\textwidth]{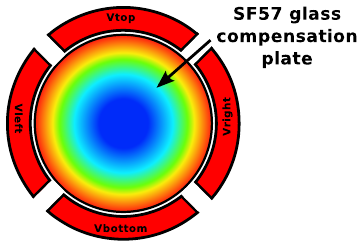}
\captionof{figure}{The design of the compensation plate. The SF57 glass is heated by four independent heaters 
to produce the required thermal profile.}
\label{fig:heater}
\end{figure}

The basic principle of the adaptive spatial mode control method is shown in 
Fig.~\ref{fig:conceptualdesign}. The Gaussian profile of the science laser beam creates 
a power-dependent thermal lens in a transmissive optic. A thermal compensation plate which can be heated from the outside 
with four individual heaters is placed after the transmissive optic. When all four heaters are heated equally, 
the defocusing thermal gradient index lens that is 
created in the compensation plate can compensate the focusing thermal lens generated in the transmissive optic. 
The use of four individual 
heaters instead of just one ring heater enables greater control over the heating profile, allowing for the correction of 
astigmatic thermal lenses and also providing beam steering capabilities.

In practice, 
the thermal compensation system will have to correct for non-stationary thermal effects, as the circulating 
laser power may change during the operation of the interferometer. To this end, the thermal compensation 
should be applied as part of a feedback loop, where any unwanted changes in the spatial beam profile are 
detected, processed, and compensated automatically. 

\section{Experimental method}
The compensator was realized in our experiment using the four segmented heater (FSH) shown in 
figure~\ref{fig:heater}, and described in Ref.~\cite{Arain10}. SF57 glass was chosen for the substrate because of its 
large thermal expansion and thermo-refractive ($\frac{dn}{dT}$) coefficients, 
which give the device a large dynamic range of achievable focal lengths. 
The four heating segments each consist of a resistive heater on Kapton film with a resistance about of 25\,$\Omega$, and are each supplied
by a current source, delivering a maximum current of 1.5\,A. This setup allows us to tune the focal length 
in the range from effectively minus infinity to -10\,m. 

\begin{figure}[h!]
\centering
\includegraphics[width=0.48\textwidth]{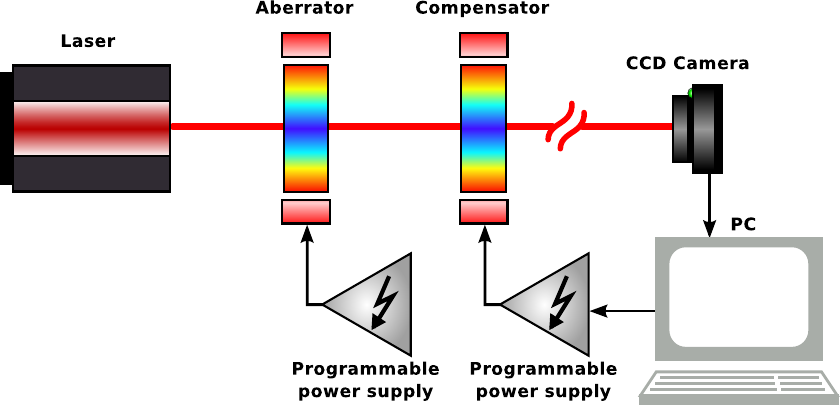}
\captionof{figure}{
A simplified experimental setup of the compensating system. Both the aberrator and compensator are 
FSHs. The aberrator was used to emulate a laser heating induced thermal 
lens. A feedback system was employed to drive the heaters in the compensator such as to maintain the original 
beam parameters, as measured at the CCD.}
\label{fig:opticslayout}
\end{figure}

A simplified experimental arrangement is shown in Fig.~\ref{fig:opticslayout}. Light from a 200\,mW Nd:YAG laser is 
passed through the first adaptive optic, referred to henceforth as the \textit{aberrator}, which is used to simulate 
the thermal lensing effect which is to be compensated. The beam transmitted through the aberrator is then 
passed through the second adaptive optic, henceforth referred to as the \textit{compensator}, which is used to 
compensate for the actions of the aberrator.
Since the laser power in the setup was not sufficient to produce a significant thermal lens, the process of thermal lensing was 
emulated in another way. Initially the aberrator was strongly biased by radially heating, such as to produce a negative focal length lens. 
A reduction in the radial heating of the aberrator from this level, and hence reduction in the power of thermal lens produced, 
was used as an analogue for an increase in central heating that would be caused by a high power laser beam. The beam 
radius at the aberrator was about 1.5\,mm (1/e$^2$ intensity). 

A Gigabit Ethernet CCD camera \cite{AlliedVision} was placed further downstream in order 
to monitor the beam transmitted through the aberrator and compensator. The 
spot size and lateral position of the beam were calculated from the CCD data, and used as a reference relative to which 
deviations could be measured and corrected. 
After processing the data, four output signals are generated; the beam radius and centroid position on the horizontal and vertical axes. 
These signals are used to control the four heating voltages on the FSH in four directions. 

In order to control 
the heating-induced thermal lens profile of the compensator accurately, it was necessary first to characterize the response of the 
compensator to actuation of each of the four heating elements. 
Differences in the mechanical contacts between the heaters and the glass, as well as between the heaters and the mount, can lead to 
different responses of the compensator to each heater. 
To map out the asymmetries in the compensator actuation, the beam first was centered 
on the CCD while all the heating elements were turned off. The voltage applied to the top heating element was steadily increased, 
and the other three voltages were adjusted in order to re-center the beam on the CCD. 
This measurement represents a relative calibration of the four different heaters, 
because unequal heating by the four heaters will lead to a change in the alignment of the transmitted beam.
The bias voltages obtained when the top heating element was supplied with 5\,V were 4.96\,V, 5.44\,V and 6.29\,V for the left, right 
and bottom heaters respectively. 

The response of the optical elements to the applied heating is far from linear over their full actuation range. 
In order to use the full range in a feedback loop it 
was therefore necessary to make a look up table, which was used to adjust the feedback filters to each heater for each set of heater 
DC offset voltages. 
Offset voltages were applied to each of the heaters in order to explore the measured beam size parameter space. These DC offsets were chosen 
such as to provide symmetric heating, as this was the actuation regime of primary interest in this experiment. 
The DC offsets required to reach specific regions within this space were recorded. At each of these points, transfer functions from applied voltage to 
measured beam parameter deviation were measured for each heater. 

\begin{figure}[htb]
\centering
\includegraphics[width=0.24\textwidth]{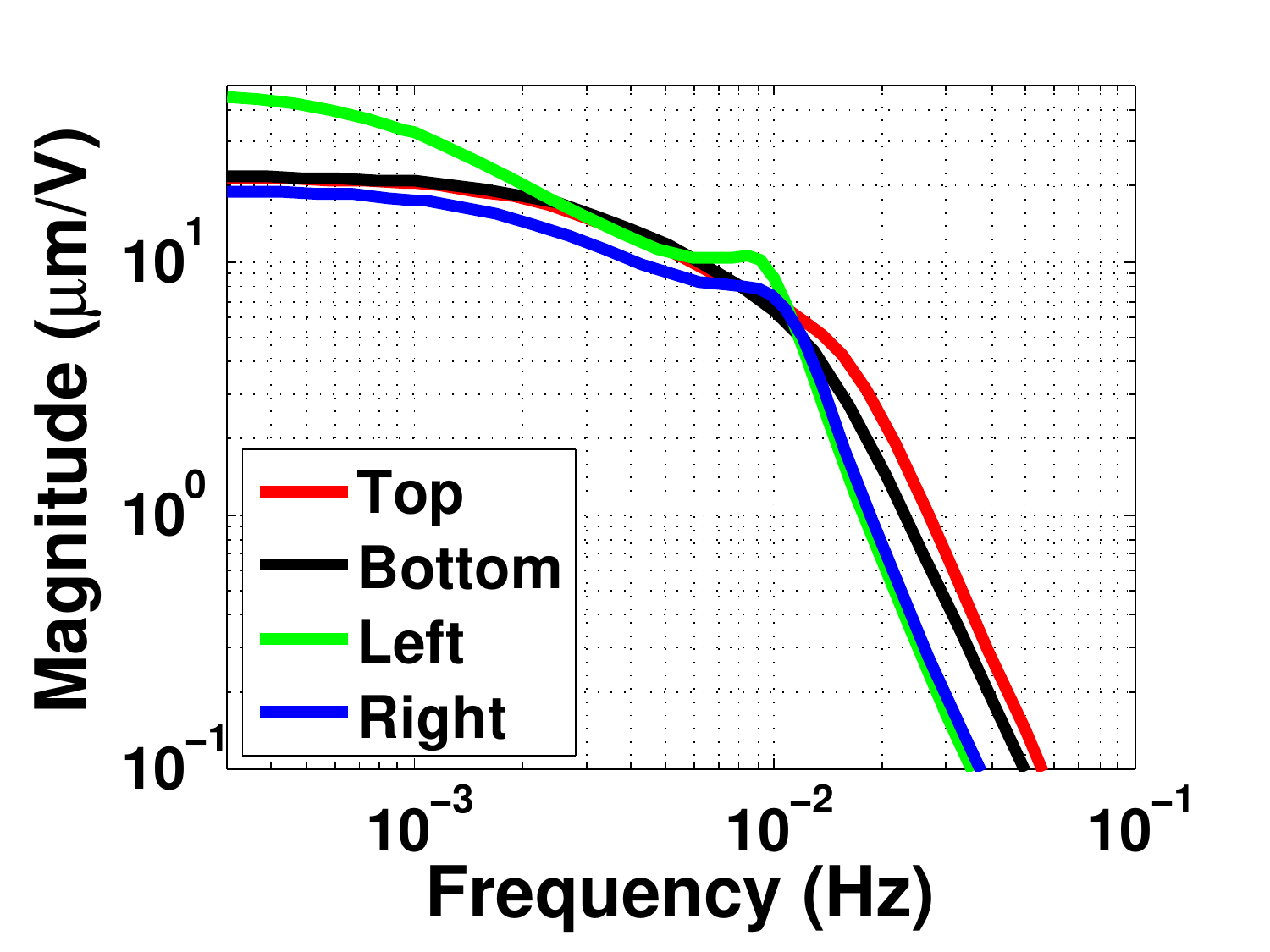}\includegraphics[width=0.24\textwidth]{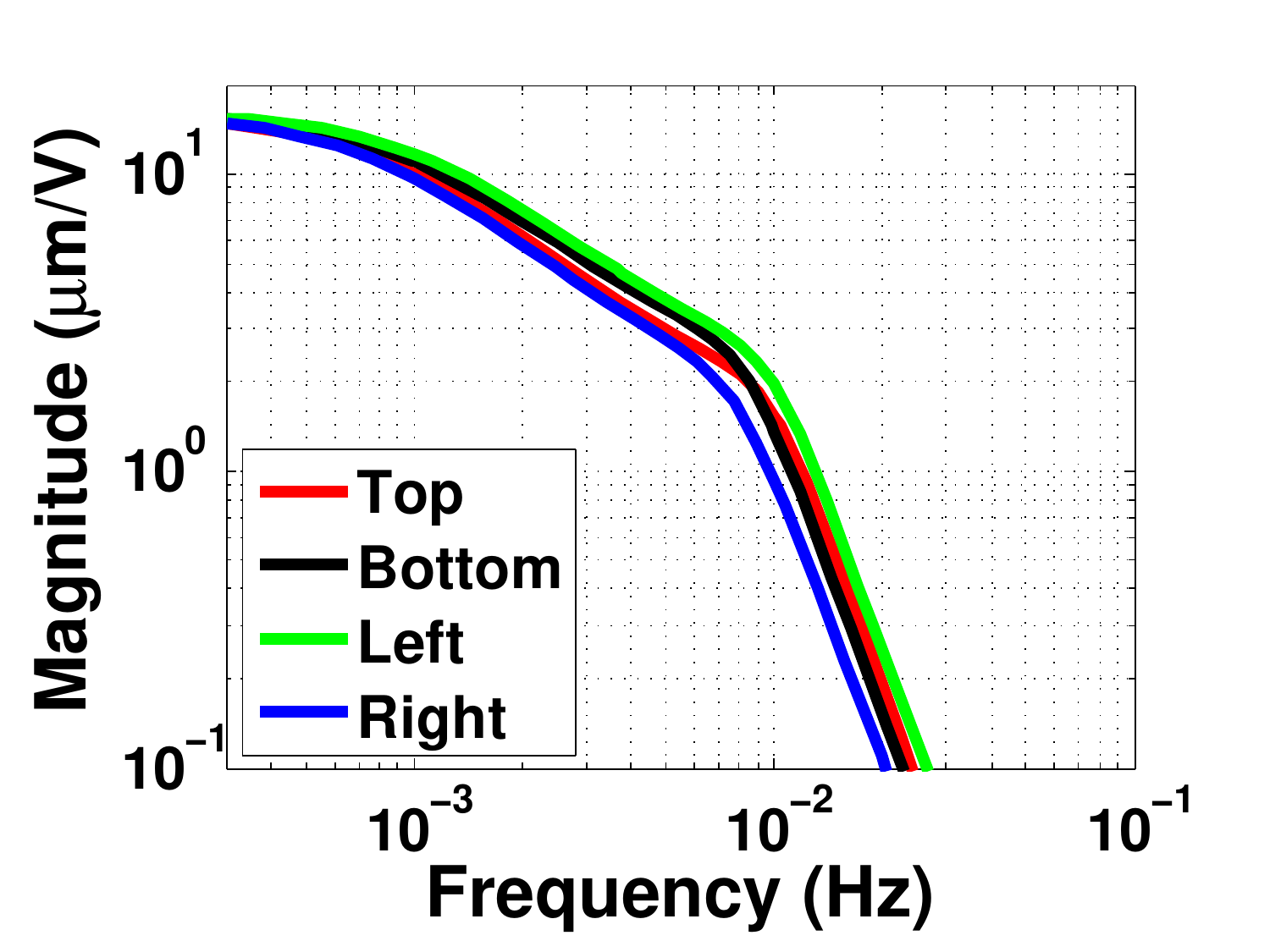}\\
\includegraphics[width=0.24\textwidth]{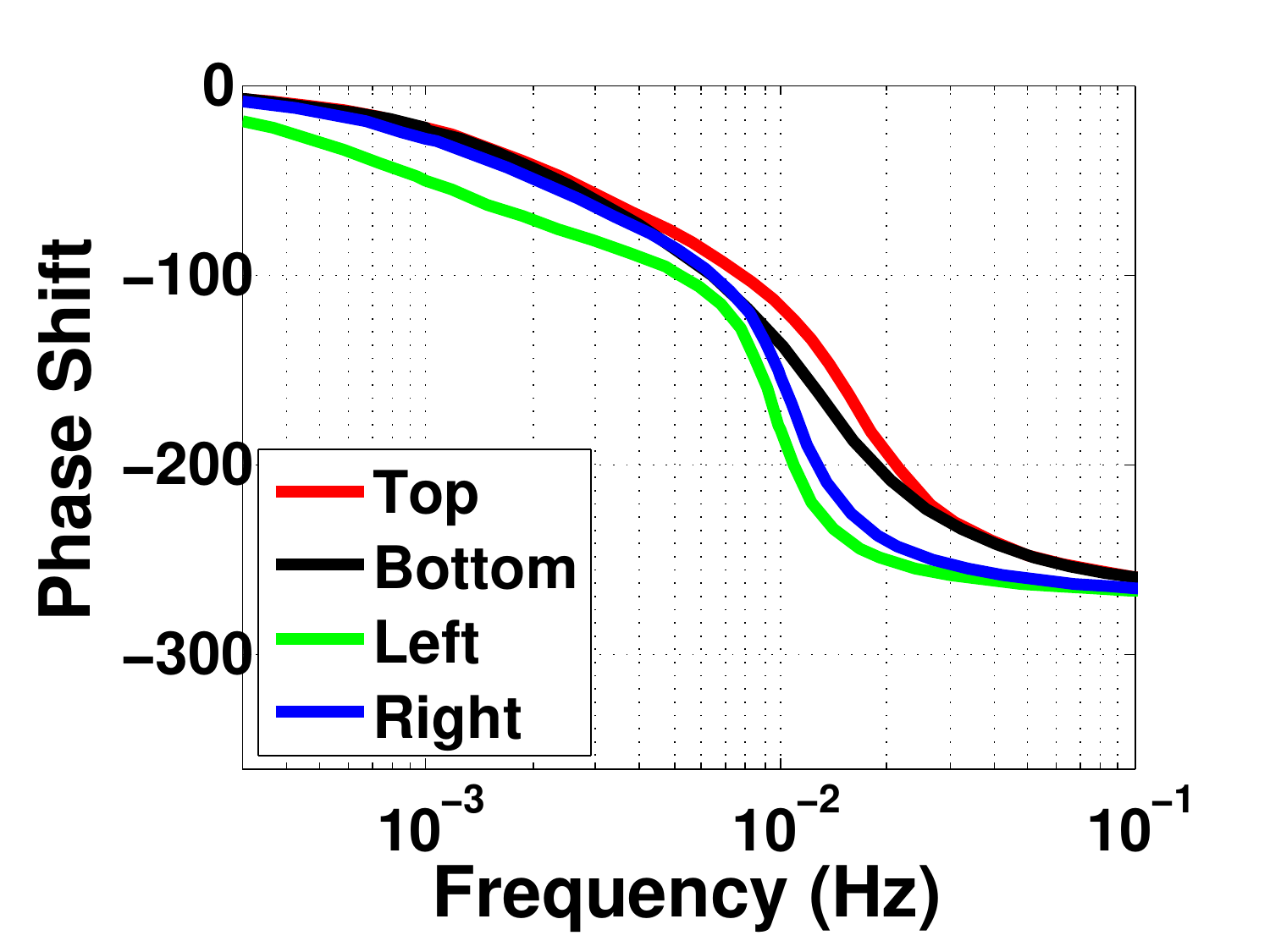}\includegraphics[width=0.24\textwidth]{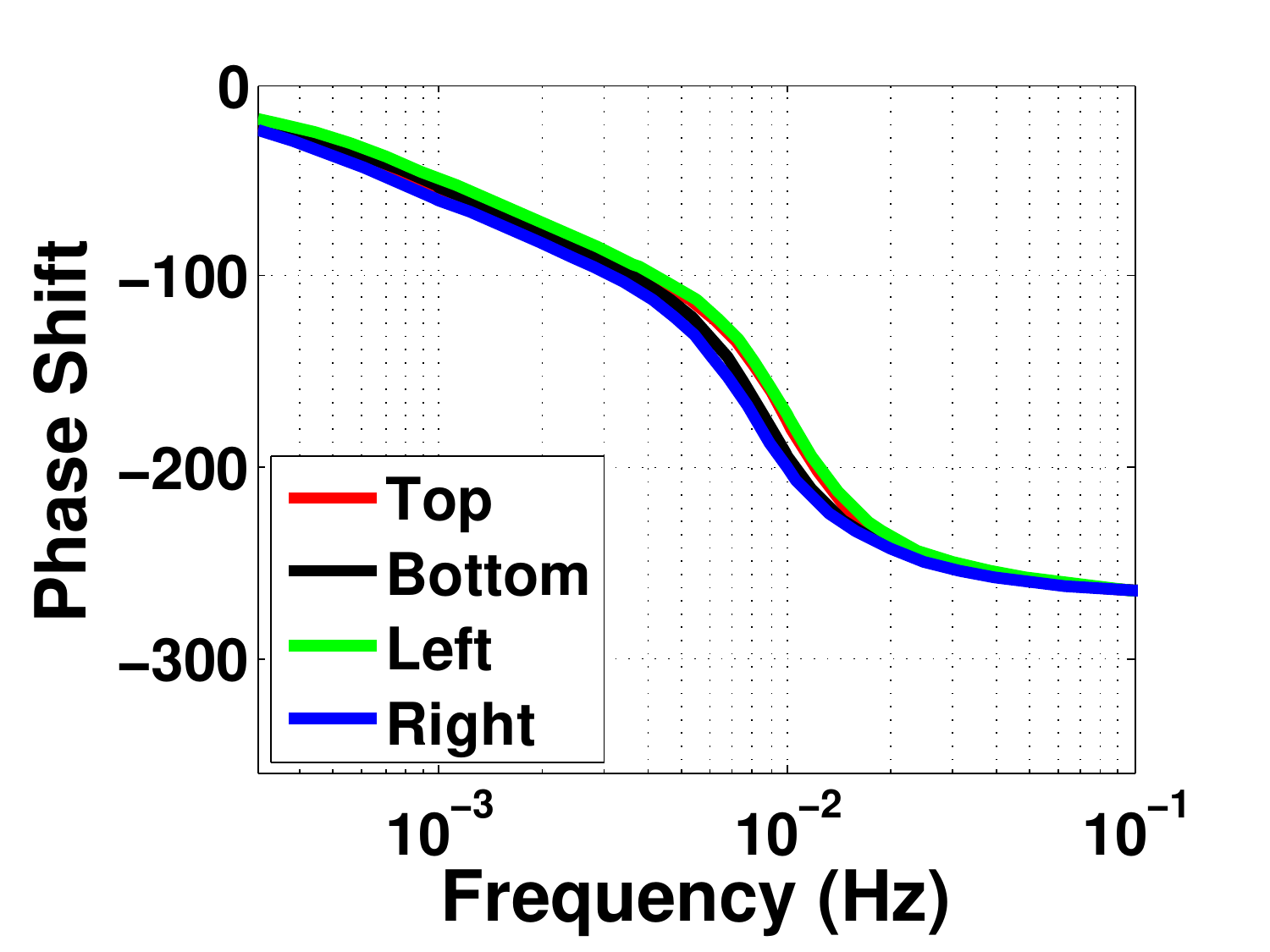}
\captionof{figure}{Transfer function of input signals $V(t)=5+0.3 \sin(2\pi ft)$\,V on the labelled heating element to beam spot width 
(left) and centroid position (right). Only the transfer functions from each actuator to its primary actuation axis are shown.}
\label{fig:TFs}
\end{figure}

The applied signal amplitude for the transfer functions was $\pm$0.3\,V; 
small enough to be within the linear range of the actuator. At 1\,mHz this linear range corresponds to a beam width 	deviation of around 7\,$\mu$m 
and a beam centroid position deviation of around 3\,$\mu$m.
Each element in the look up table was then composed from the DC bias offset values 
required to bring the beam parameters \emph{close} to the working point, and the transfer functions from applied signal to beam parameter deviation 
within the linear range in order to bring the beam parameters to the precise working point. Figure \ref{fig:TFs} shows the fitted 3-pole transfer function 
from each heater to the corresponding beam parameters for one set of DC voltage values from the look up table.

These transfer functions were used to describe the system in term of its transfer matrix:
\begin{equation}
\begin{bmatrix}
\Delta w_\mathrm{H} \\
\Delta p_\mathrm{H} \\
\Delta w_\mathrm{V} \\
\Delta p_\mathrm{V}
\end{bmatrix}
=
\begin{bmatrix}
M_{11}&M_{12}&M_{13}&M_{14}\\
M_{21}&M_{22}&M_{23}&M_{24}\\
M_{31}&M_{32}&M_{33}&M_{34}\\
M_{41}&M_{42}&M_{43}&M_{44}
\end{bmatrix}
\begin{bmatrix*}[l]
\Delta V_\mathrm{left} \\
\Delta V_\mathrm{right} \\
\Delta V_\mathrm{top} \\
\Delta V_\mathrm{bottom}
\end{bmatrix*},
\label{eqn:VBSP}\end{equation}
where $w$ and $p$ represent beam width and position respectively, and the subscripts H and V represent horizontal 
and vertical directions respectively; $\Delta V$ are the small 
actuation voltages (not the bias voltages) applied to the left, right, top and and bottom sections of the FSH; and $M$ is the 4$\times$4 transfer matrix.  
The control matrix for feedback to the compensator was obtained by inverting this transfer matrix. 

\section{Experimental demonstration}
\subsection{Compensation of constant uniform thermal aberration}
\begin{figure}[htb]
\centering
\includegraphics[width=0.175\textwidth]{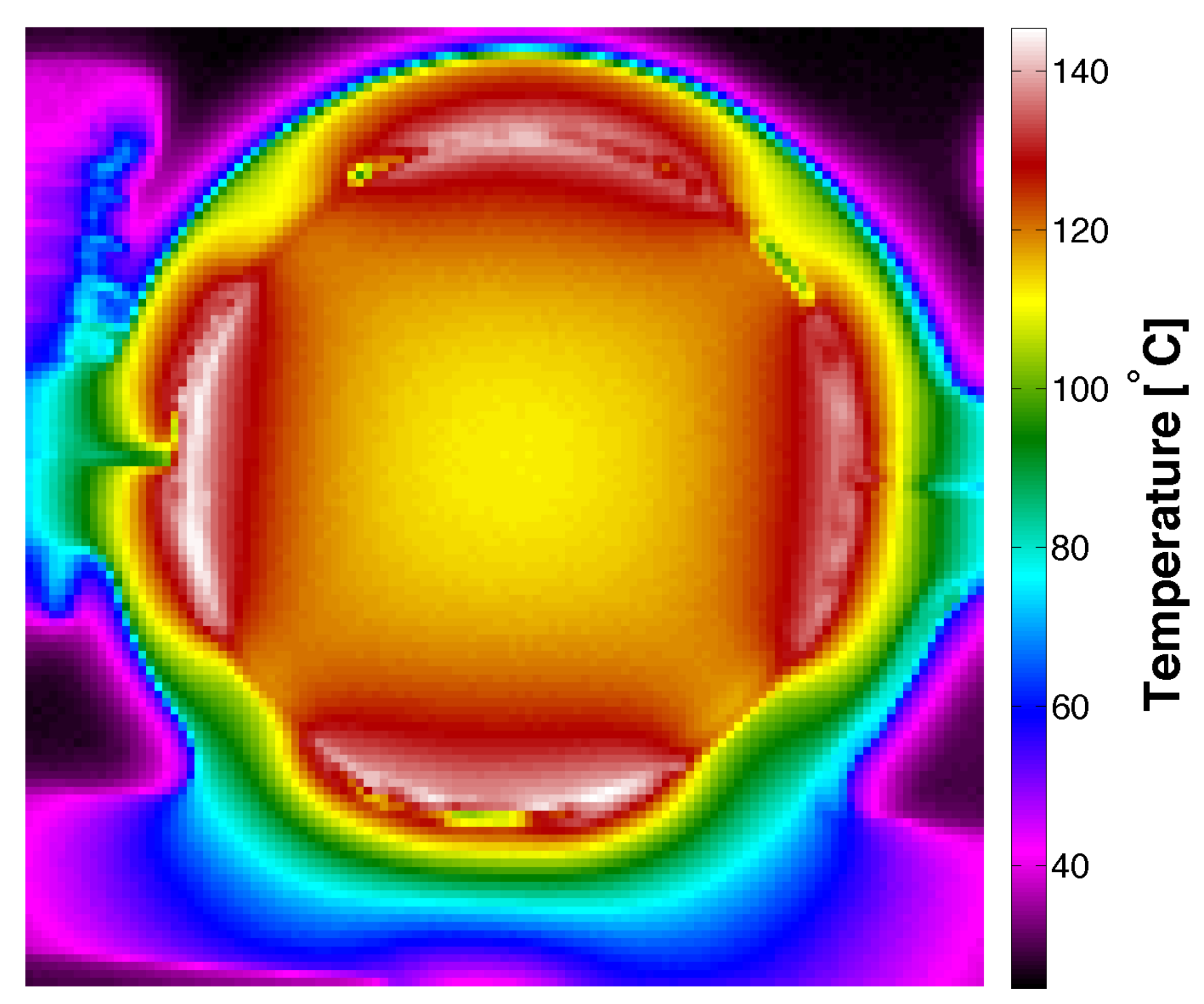}
\hspace{0.5cm}
\includegraphics[width=0.175\textwidth]{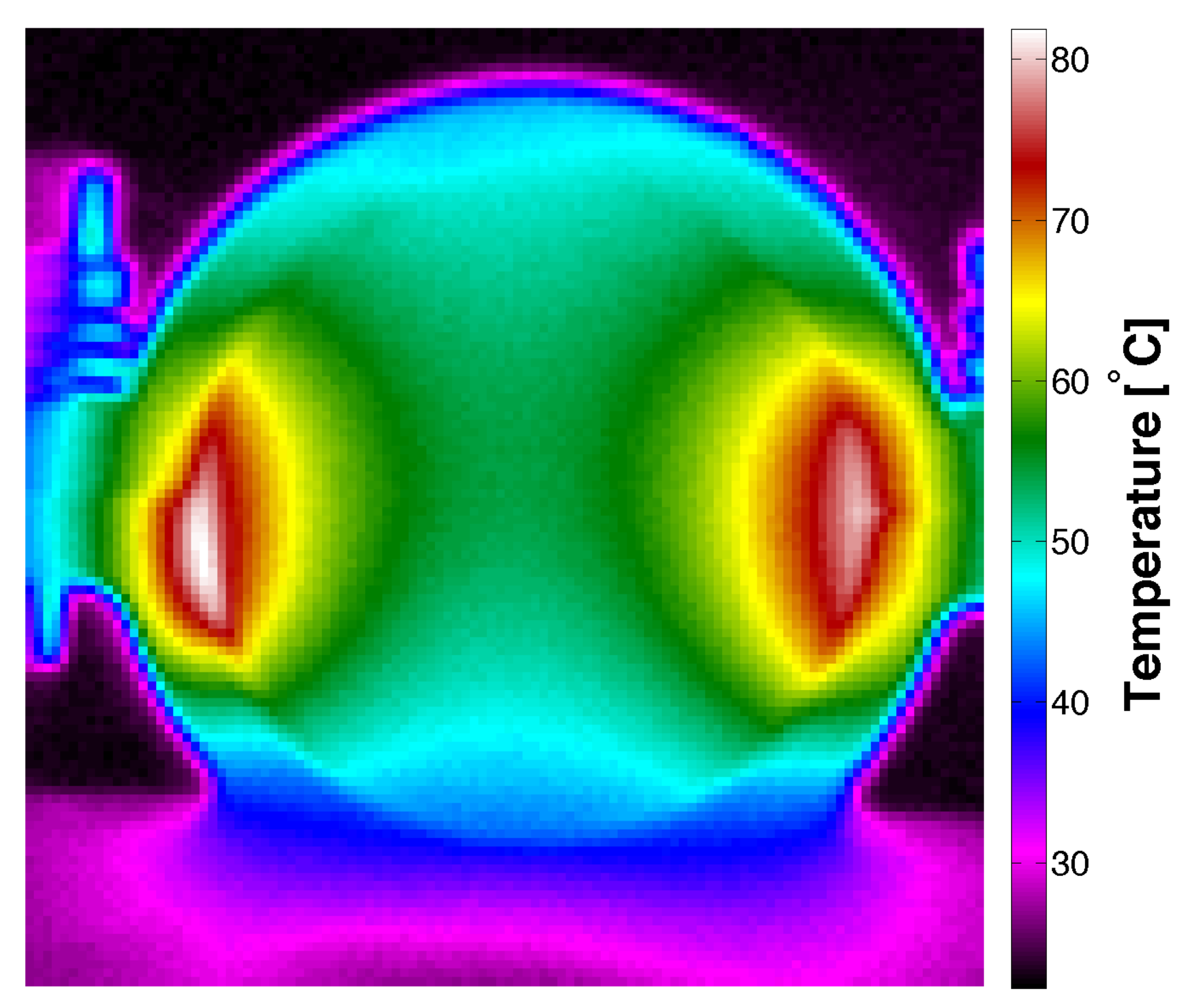}
\caption{The temperature profile of the aberrator under the symmetric (left) and astigmatic (right) heating conditions.}
\label{fig:Timages}
\end{figure}

In this section we demonstrate feedback compensation for circularly symmetric thermal aberration, such as may be caused by absorption of high laser 
powers in transmissive optics under normal incidence. The aberrator was uniformly heated at each quadrant until thermal equilibrium was reached, 
and then the heating power was turned off. The left panel of Fig. \ref{fig:Timages} shows the temperature profile of the aberrator at thermal equilibrium; 
in the central part of the optic the induced temperature gradient creates a good approximation to a spherical thermal lens.

\begin{figure}[htb]
\centering
\includegraphics[width=0.48\textwidth]{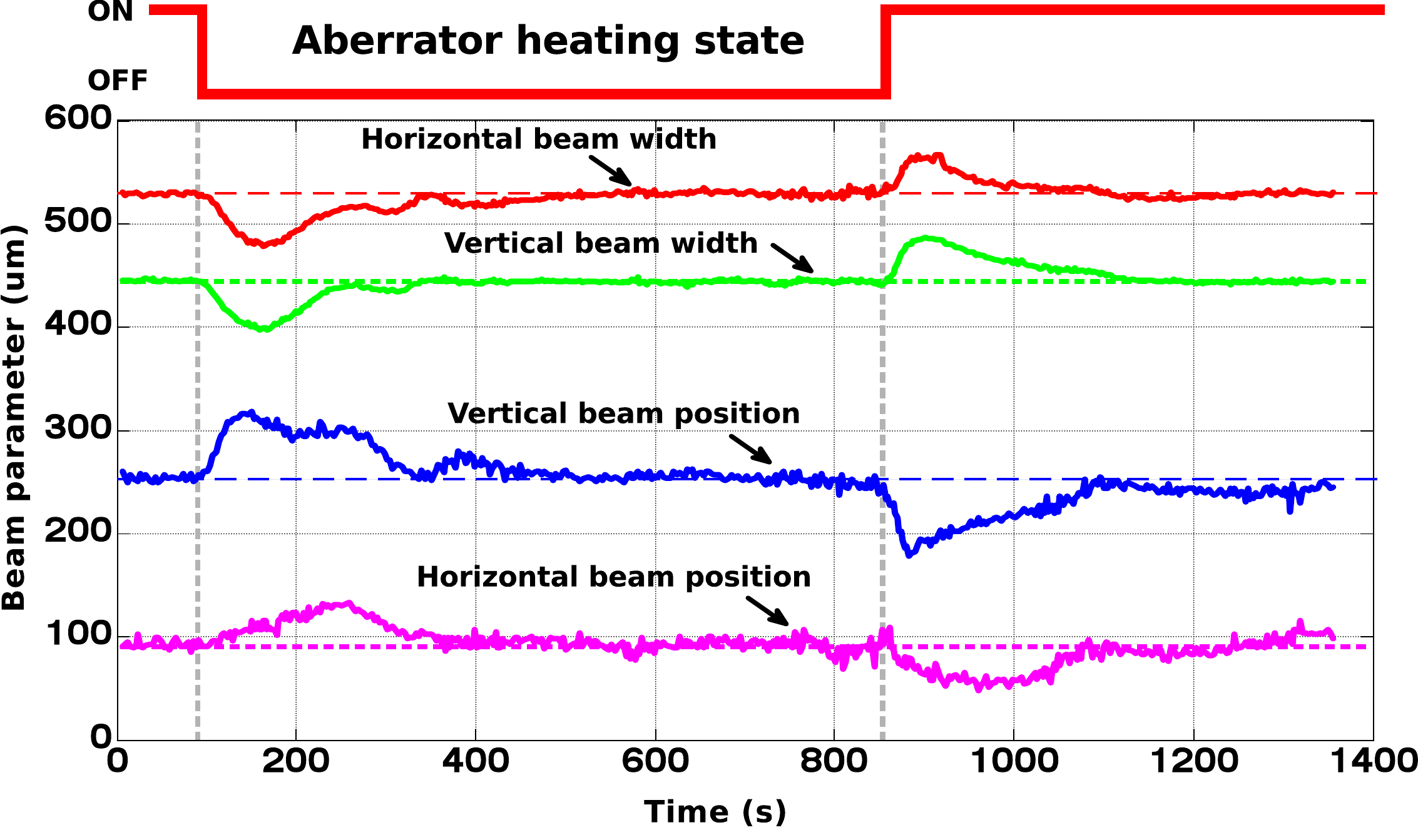}
\captionof{figure}{Time series of the measured beam parameters while the beam-shaping feedback loop was closed. Impulsive beam aberrations were applied at 
90\,s and 860\,s.}
\label{fig:inloopmeasurements}
\end{figure}

Figure \ref{fig:inloopmeasurements} shows time domain measurements of all 4 of the measured beam parameters while the beam shaping feedback loop was closed. 
The horizontal and vertical widths of the beam are different since the available laser output beam is elliptical, and no attempt was made to circularize it.
During the run, two impulsive aberrations were generated --- one at 90\,s (symmetrically reducing the aberrator heating) 
and one at 860\,s (re-applying the aberrator heating). After each impulsive aberration, the parameters recovered to their set-point values within 
approximately 300\,s.

Initially, as the aberrator cools down and its effective 
focal length becomes less negative, the beam parameters measured at the CCD begin to change. The beam sizes in both axes decrease at first, as expected 
due to the change in focal length. The beam spot position in both axes also begins to change; this change may be caused either by imperfect 
centering of the beam on the aberrator, or by differences in the rate of heat loss from different areas of the aberrator optic. For example, one may expect 
a greater rate of heat loss through conduction in the the lower portion of the aberrator due to its mechanical contact with the steel post on which it is mounted. 
This will lead to a linear component in the thermal gradient across the optic, thus causing a shift in the transmitted beam position.

As the beam parameters begin to change, a slow integrator in the feedback signal path determines and applies the 
bias voltages required to return the beam parameters close to the set-points. During this time a linear proportional-integral controller 
also applies small signals (within the linear regime), filtered by the transfer matrix elements corresponding to the closest current DC offset voltage 
look up table elements, to compensate for small deviations in each of the four beam parameters around the working point. The beam 
parameters return to the set-point values by around 300\,s after the impulsive aberration; 
the compensation system has regained the starting beam parameters at the CCD, despite the continued presence of the aberration. 

\subsection{Compensation of astigmatic thermal aberration}\label{subsec:astig}
In this section we demonstrate the ability of the compensator to correct for astigmatic heating by applying heat in the vertical or horizontal axes only. 
This experiment was similar to the axisymmetric heating experiment, except that the heating was only applied to the aberrator across the horizontal axis, 
at the left and right heaters. In the initial state the aberrator was heated astigmatically until it reached thermal equilibrium, producing the temperature 
profile shown in the right panel of Fig.~\ref{fig:Timages}. Three impulsive astigmatic aberrations were then applied by stopping the heating after 420\,s, 
reapplying the heating after 1800\,s, and finally ceasing the heating after 2450\,s. 
\begin{figure}[htb]
\centering
\includegraphics[width=0.48\textwidth]{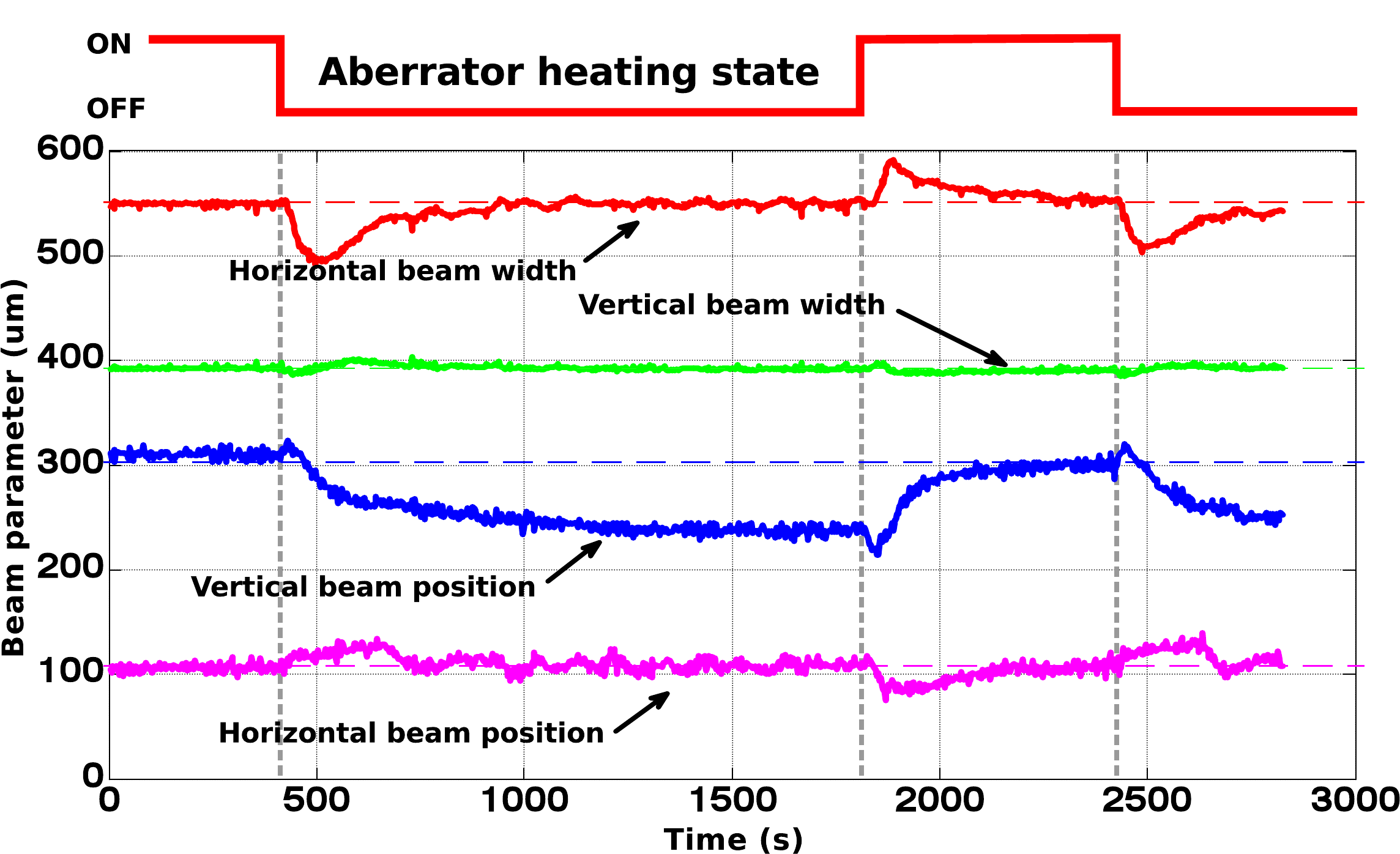}
\captionof{figure}{Time series of the measured beam parameters while the beam-shaping feedback loop was closed, for astigmatic aberrations. 
Impulsive astigmatic beam aberrations were applied at 420\,s, 1800\,s and 2450\,s.}
\label{fig:astigmaticcomp}
\end{figure}

Figure~\ref{fig:astigmaticcomp} shows 
the time series of the four measured beam parameters throughout the measurement period. As in Fig.~\ref{fig:inloopmeasurements}, it can be seen 
that the control loop returns the beam sizes measured at the CCD to their initial set-point values before the impulsive aberrations were applied. 
However, it is also clear that the beam position along the vertical axis was not well controlled by the feedback loop. When the aberration is first 
applied the beam drifts lower, and this error is not corrected for by the compensator. 

This drift is most likely caused by the fact that the look up table was only measured for symmetric DC offset values, i.e. those used for symmetric 
compensation. As a result, it was decided to weight the feedback to beam size more heavily than the feedback to beam position for 
the astigmatic compensation case. 
If necessary, a separate look up table could be made for purely astigmatic compensation, and one may consider interpolating between 
this and the symmetric compensation look up table in order to cover the full range of compensation. In a realistic application of this device, however, 
alignment drifts such as those observed here can be compensated easily using movable mirrors. 

\subsection{Noise analysis}
As with any feedback loop, it is important to ascertain the level of noise injected into the system as a consequence of implementing 
the loop. To this end, the standard deviations in beam parameters were calculated with and without actuation on the aberrator. 
In the state where neither the aberrator nor compensator were heated, the standard deviations in beam radius were 0.15\,\% and 0.16\,\% 
of the mean beam radius in the horizontal and vertical axes respectively, and the standard deviations in beam pointing angle were 
0.27\,$\mu$rad and 0.57\,$\mu$rad for the horizontal and vertical axes. In the case where the aberrator was heated with the strong bias voltage, 
the corresponding standard deviations increased to 0.40\,\%, 0.33\,\% and 2.9\,$\mu$rad and 2.1\,$\mu$rad. Here it can be seen that even static
actuation of the FSHs leads to a significant increase in beam parameter noise. This is almost certainly a consequence of operating the 
devices in air; the high temperatures at the element (>100$^\circ$C) cause turbulent air flow in its vicinity resulting in increased jitter downstream. 
This increased jitter is expected to be strongly mitigated when the actuators are used in vacuum.

\section{Conclusions}
It has been demonstrated that a segmented heating-compensating system can be used as an actuator in an active feedback loop to correct for 
time-dependent thermally-induced aberrations of optics. The compensation of both uniform and astigmatic aberrations was tested. The uniform 
aberration compensation was shown to return all measured beam parameters to their initial values within around 300\,s of the application of 
an impulsive thermal aberration. The astigmatic aberration compensation showed a similar performance for most of the measured beam parameters, 
although it was unable to compensate for a drift in the position of the beam in a direction orthogonal to the axis of compensation. This limitation 
could likely be improved by better beam centering on the optics or by implementing a look up table made specifically for astigmatic compensation, 
or could be corrected for by the use of steering mirrors in the beam path. 

The use of the actuators was shown to increase the standard deviation of the measured beam parameters, though this is almost certainly 
due to the effects of turbulent air flow around the FSHs as a result of their high temperature relative to the environment.
These actuators are primarily designed for in-vacuum operation, however, so the turbulence effect does not present a severe limitation on 
their usefulness. Further studies into the vacuum operation of the actuators, and the impact 
of the actuators on the quality of the beam transmitted through them are recommended in order to determine the potential for this device to make an 
impact in achieving the high-power operation of aLIGO. In the future, it may also be interesting to investigate the application of the device presented here 
as an actuator in a control loop to stabilize mode matching into an interferometer, similar to that described in Ref.~\cite{Mueller00}.

\section{Acknowledgments}
The authors acknowledge the encouragement of the LIGO Science Collaboration. Prof. Prabir Barooah is also thanked for his 
helpful discussions. This work is supported by the National Science Foundation under grant under grants PHY-0855313 and and PHY-12055FSH12.
This paper has been assigned the LIGO document No. LIGO-P1300045.

\end{document}